\documentclass[aps, twocolumn, superscriptaddress, showpacs]{revtex4-1}

\pdfoutput=1
\usepackage{graphicx}
\usepackage{amssymb,amsmath}
\usepackage{bm}

\newcommand{\la}[0]{{\Big\langle}}
\newcommand{\ra}[0]{{\Big\rangle}}

\begin{document}

\title{Optimization of linear and nonlinear interaction schemes for stable synchronization of weakly coupled limit-cycle oscillators}

\author{Nobuhiro Watanabe}
\affiliation{Department of Systems and Control Engineering,
Tokyo Institute of Technology, Tokyo 152-8552, Japan}

\author{Yuzuru Kato}
\email{Corresponding author: kato.y.bg@m.titech.ac.jp}
\affiliation{Department of Systems and Control Engineering,
Tokyo Institute of Technology, Tokyo 152-8552, Japan}

\author{Sho Shirasaka}
\affiliation{Department of Information and Physical Sciences,
Graduate School of Information Science and Technology,
Osaka University
1-5 Yamadaoka, Suita, Osaka 565-0871, Japan
}

\author{Hiroya Nakao}
\affiliation{Department of Systems and Control Engineering,
Tokyo Institute of Technology, Tokyo 152-8552, Japan}

\begin{abstract}
Optimization of mutual synchronization between a pair of limit-cycle oscillators with weak symmetric coupling is considered in the framework of the phase reduction theory. By generalizing a previous study~\cite{shirasaka17b} on the optimization of cross-diffusion coupling matrices between the oscillators, we consider optimization of mutual coupling signals to maximize the linear stability of the synchronized state, which are functionals of the past time sequences of the oscillator states. For the case of linear coupling, optimization of the delay time and linear filtering of coupling signals are considered. For the case of nonlinear coupling, general drive-response coupling is considered, and the optimal response and driving functions are derived. The theoretical results are exemplified by numerical simulations.
\end{abstract}


\maketitle

\section{Introduction}

Synchronization of rhythmic dynamical elements exhibiting periodic oscillations is widely observed in the real world~\cite{winfree80,pikovsky01,strogatz03}. Collective oscillations arising from the mutual synchronization of dynamical elements often play important functional roles, such as the synchronized secretion of insulin from pancreatic beta cells~\cite{winfree80,sherman91} and synchronized oscillation of power generators~\cite{strogatz03,motter13,dorfler}. Clarifying the mechanisms of collective synchronization and devising efficient methods of mutual synchronization are thus both fundamentally and practically important.

The stable periodic dynamics of rhythmic elements are often modeled as limit-cycle oscillators~\cite{winfree80,pikovsky01,strogatz03,strogatz15}.
When mutual interactions between limit-cycle oscillators are weak, synchronization dynamics of the oscillators can be analyzed using the phase reduction theory~\cite{winfree80,kuramoto84,hoppensteadt97,ermentrout10,nakao16,ashwin16}.
In this approach, nonlinear multi-dimensional dynamics of an oscillator is reduced to a simple approximate phase equation, characterized by the natural frequency and phase sensitivity of the oscillator.
The phase reduction theory facilitates systematic and detailed analysis of synchronization dynamics. It has been used to explain nontrivial synchronization dynamics of 
coupled oscillators, such as the collective synchronization transition of an ensemble of coupled oscillators~\cite{winfree80,kuramoto84,hoppensteadt97,ermentrout10,nakao16,ashwin16}.
Generalization of the method for non-conventional limit-cycling systems, such as time-delayed oscillators~\cite{novicenko12,kotani12}, hybrid oscillators~\cite{shirasaka17a,park18}, collectively oscillating networks~\cite{kawamura08}, and rhythmic spatiotemporal patterns~\cite{kawamura13,nakao14}, has also been discussed.

Recently, the phase reduction theory has been applied for the control and optimization of synchronization dynamics in oscillatory systems.
For example, Moehlis {\it et al.}~\cite{moehlis06}, Harada {\it et al.}~\cite{harada10}, Dasanayake and Li~\cite{dasanayake11}, Zlotnik {\it et al.}~\cite{zlotnik12,zlotnik13,zlotnik16}, Pikovsky~\cite{pikovsky15}, Tanaka {\it et al.}~\cite{tanaka14a,tanaka14b,tanaka15}, Wilson {\it et al.}~\cite{wilson2015}, Pyragas {\it et al.}~\cite{pyragas2018}, and Monga {\it et al.}~\cite{monga2018a,monga2018b} have used the phase reduction theory (as well as the phase-amplitude reduction theory) to derive optimal driving signals for the stable entrainment of nonlinear oscillators in various physical situations.

In a similar spirit, in our previous study~\cite{shirasaka17b}, we considered a problem of improving the linear stability of synchronized state between a pair of limit-cycle oscillators by optimizing a cross-diffusion coupling matrix between the oscillators, where different components of the oscillators are allowed to interact. We also considered a pair of mutually interacting reaction-diffusion systems exhibiting rhythmic spatiotemporal patterns and derived optimal spatial filters for stable mutual synchronization~\cite{kawamura17}.

In this study, we consider this problem in a more general setting, whereby the oscillators can interact not only by their present states but also through the time sequences of their past states.
We first consider linear coupling with time delay or temporal filtering and derive the optimal delay time or linear filter.
We then consider general nonlinear coupling with a mutual drive-response configuration and derive the optimal response function and driving function for stable synchronization. We argue that, although we consider general coupling that can depend on the past time sequences of the oscillators, the optimal mutual coupling can be obtained as a function of the present phase values of the oscillators in the framework of the phase-reduction approximation.
The results are illustrated by numerical simulations using Stuart-Landau and FitzHugh-Nagumo oscillators as examples.

This paper is organized as follows. In Sec. II, we introduce a general model of coupled limit-cycle oscillators and reduce it to coupled phase equations. In Sec. III, we consider the case with linear coupling and derive the optimal time delay and optimal linear filter for coupling signals. In Sec. IV, we consider nonlinear coupling of the drive-response type and derive the optimal response function and driving function. In Sec. V, a summary is provided.

\section{Model}

\subsection{Pair of weakly coupled oscillators}

In this study, we consider a pair of weakly and symmetrically coupled limit-cycle oscillators with identical properties, where the oscillators can mutually interact not only through their present states but also through their past time sequences.
We assume that the oscillators are generally described by the following equations:
\begin{align}
\label{eq1}
\dot{\bm X}_1(t) &= {\bm F}({\bm X}_1(t)) + \epsilon \hat{\bm H}\{ {\bm X}_1^{(t)}(\cdot), {\bm X}_2^{(t)}(\cdot) \},
\cr
\dot{\bm X}_2(t) &= {\bm F}({\bm X}_2(t)) + \epsilon \hat{\bm H}\{ {\bm X}_2^{(t)}(\cdot), {\bm X}_1^{(t)}(\cdot) \},
\end{align}
where ${\bm X}_{1, 2} \in {\mathbb R}^N$ are $N$-dimensional state vectors of the oscillators $1$ and $2$ at time $t$, ${\bm F} : {\mathbb R}^N \to {\mathbb R}^N$ is a sufficiently smooth vector field representing the dynamics of individual oscillators,
and 
$\epsilon \hat{\bm H}$ represents weak mutual coupling between the oscillators.
Here, $\hat{\bm H} : C \times C \to {\mathbb R}^{N}$ ($C$ is a function space of the time sequences of length $T$) is a functional of two vector functions; i.e., the past time sequences of ${\bm X}_{1, 2}(t)$, and $0 < \epsilon \ll 1$ is a small parameter representing the smallness of the mutual coupling.

We assume that an isolated system, $\dot{\bm X} = {\bm F}({\bm X})$, has an exponentially stable limit cycle $\tilde{\bm X}_0(t) = \tilde{\bm X}_0(t+T)$ of period $T$ and frequency $\omega = 2\pi / T$, and the system state deviates only slightly from this limit cycle if weak perturbations are applied.
The time sequences ${\bm X}_{i}^{(t)}(\cdot)$ ($i=1,2$) in the coupling functional $\hat{\bm H}$ represent functions ${\bm X}_{i}(\tau)$ on the interval $[t-T, t]$ and are defined as
\begin{align}
{\bm X}_{i}^{(t)}(\tau) = {\bm X}_{i}(t + \tau) \quad (-T \leq \tau \leq 0).
\end{align}
In this study, we fix the length of the time sequences used for the coupling as the natural period $T$ of the limit-cycle oscillator.

Under the assumptions stated above, we can employ the standard method of phase reduction~\cite{winfree80,kuramoto84,hoppensteadt97,ermentrout10,nakao16,ashwin16} and introduce a phase function $\Theta({\bm X}) : {\mathbb R}^N \to [0, 2\pi)$, which satisfies $\dot\Theta({\bm X}) = {\bm F}({\bm X}) \cdot \nabla \Theta({\bm X}) = \omega$ in the basin of the limit cycle. Using $\Theta({\bm X})$, the phase variable of this oscillator can be defined as $\theta = \Theta( {\bm X})$, which constantly increases with time as $\dot{\theta} = \omega$, both on and in the basin of the limit cycle ($2\pi$ is identified with $0$ as usual).
The oscillator state on the limit cycle can be represented as a function of $\theta$ as ${\bm X}_0(\theta) = \tilde{\bm X}_0(t = \theta / \omega)$, which is a $2\pi$-periodic function of $\theta$, ${\bm X}_0(\theta) = {\bm X}_0(\theta+2\pi)$.
Similarly, to represent a time sequence on the limit cycle, we also introduce the notation
\begin{align}
{\bm X}_0^{(\theta)}(\tau) = {\bm X}_0(\theta + \omega \tau) \quad (- T \leq \tau \leq 0)
\end{align}
and abbreviate this as ${\bm X}_0^{(\theta)}(\cdot)$.
We hereafter use these notations to represent the system states and their time sequences on the limit cycle.

\subsection{Phase reduction and averaging}

Assuming that perturbations applied to the oscillator are sufficiently weak, the state vector of the oscillator can be represented approximately by ${\bm X}(t) \approx {\bm X}_0(\theta(t))$ as a function of the phase $\theta(t)$.
Then, the phase sensitivity function (PSF), defined by ${\bm Z}(\theta) = \nabla \Theta({\bm X})|_{{\bm X} = {\bm X}_0(\theta)} : [0, 2\pi) \to {\mathbb R}^N$, characterizes the linear response property of the oscillator phase to weak perturbations.
When the oscillator is weakly driven by a perturbation, ${\bm p}$, as $\dot{\bm X} = {\bm F}({\bm X}) + \epsilon {\bm p}$, the reduced phase equation is given by $\dot{\theta} = \omega + \epsilon {\bm Z}(\theta) \cdot {\bm p}$ up to $O(\epsilon)$.
The PSF can be calculated as a $2\pi$-periodic solution to an adjoint equation $\omega d{\bm Z}(\theta)/d\theta = - {\rm J}^{\dag}({\bm X}_0(\theta)) {\bm Z}(\theta)$, with a normalization condition ${\bm Z}(\theta) \cdot d{\bm X_0(\theta)}/{d\theta} = 1$, where ${\rm J}({\bm X}) : {\mathbb R}^N \to {\mathbb R}^{N \times N}$ is a Jacobian matrix of ${\bm F}$ at ${\bm X} = {\bm X}_0(\theta)$ and $\dag$ denotes transpose.
In the numerical analysis, ${\bm Z}(\theta)$ can be calculated easily using the backward time-evolution of the adjoint equation as proposed by Ermentrout~\cite{ermentrout10}.

Defining the phase values of the oscillators $1, 2$ as $\theta_{1, 2} = \Theta({\bm X}_{1, 2})$,
the following pair of equations can be derived from Eq.~(\ref{eq1}):
\begin{align}
\dot{\theta}_1(t) &= \omega + \epsilon {\bm Z}(\theta_1(t)) \cdot \hat{\bm H}\{ {\bm X}_1^{(t)}(\cdot), {\bm X}_2^{(t)}(\cdot) \},
\cr
\dot{\theta}_2(t) &= \omega + \epsilon {\bm Z}(\theta_2(t)) \cdot \hat{\bm H}\{ {\bm X}_2^{(t)}(\cdot), {\bm X}_1^{(t)}(\cdot) \},
\label{phase00}
\end{align}
which are correct up to $O(\epsilon)$.
Next, we use the fact that the deviation of each oscillator state from the limit cycle is small and of $O(\epsilon)$:
\begin{align}
{\bm X}_{1,2}^{(t)}(\tau) = {\bm X}_{0}^{(\theta_{1,2}(t))}(\tau) + O(\epsilon)
\quad
(-T \leq \tau \leq 0).
\end{align}
Substituting into Eq.~(\ref{phase00}) and ignoring errors of $O(\epsilon^2)$, we obtain a pair of reduced phase equations,
\begin{align}
\dot{\theta}_1 &= \omega + \epsilon {\bm Z}(\theta_1) \cdot \hat{\bm H}\{ {\bm X}_0^{(\theta_1)}(\cdot), {\bm X}_0^{(\theta_2)}(\cdot) \},
\cr
\dot{\theta}_2 &= \omega + \epsilon {\bm Z}(\theta_2) \cdot \hat{\bm H}\{ {\bm X}_0^{(\theta_2)}(\cdot), {\bm X}_0^{(\theta_1)}(\cdot) \},
\end{align}
which are also correct up to $O(\epsilon)$.

The coupling term, $\hat{\bm H}\{ {\bm X}_0^{(\theta_1)}(\cdot), {\bm X}_0^{(\theta_2)}(\cdot) \}$, is a functional of the two time sequences ${\bm X}_0^{(\theta_1)}(\cdot)$ and ${\bm X}_0^{(\theta_2)}(\cdot)$.
However, because we can neglect the deviations of the oscillator states from the limit cycle at the lowest order approximation, the functionals of ${\bm X}_0^{(\theta_1)}(\cdot)$ and ${\bm X}_0^{(\theta_2)}(\cdot)$ are actually determined solely by the two phase values $\theta_1$ and $\theta_2$.
Therefore, we can regard the coupling term as an ordinary function of $\theta_1$ and $\theta_2$, defined by
\begin{align}
{\bm H}(\theta_1, \theta_2) = \hat{\bm H}\{ {\bm X}_0^{(\theta_1)}(\cdot), {\bm X}_0^{(\theta_2)}(\cdot) \},
\end{align}
and rewrite the phase equations as
\begin{align}
\label{eqphase}
\dot{\theta}_1 &= \omega + \epsilon {\bm Z}(\theta_1) \cdot {\bm H}( \theta_1, \theta_2 ),
\cr
\dot{\theta}_2 &= \omega + \epsilon {\bm Z}(\theta_2) \cdot {\bm H}( \theta_2, \theta_1 ).
\end{align}
Thus, though we started from Eq.~(\ref{eq1}) with general coupling functionals that depend on the past time sequences of the oscillators,
the coupled system reduces to a pair of simple ordinary differential equations that depend only on the present phase values $\theta_1$ and $\theta_2$ of the oscillators within the phase-reduction approximation.

Following the standard averaging procedure~\cite{kuramoto84,hoppensteadt97}, we introduce slow phase variables $\phi_{1, 2} = \theta_{1, 2} - \omega t$, rewrite the equations as
\begin{align}
\dot{\phi}_1 &= \epsilon {\bm Z}(\phi_1+\omega t) \cdot {\bm H}( \phi_1+\omega t, \phi_2+\omega t ),
\cr
\dot{\phi}_2 &= \epsilon {\bm Z}(\phi_2+\omega t) \cdot {\bm H}( \phi_2+\omega t, \phi_1+\omega t ),
\end{align}
and average the small right-hand side of these equations over one-period of oscillation.
This yields the following averaged phase equations, which are correct up to $O(\epsilon)$:
\begin{align}
\label{eq7}
\dot{\theta}_1 &= \omega + \epsilon \Gamma( \theta_1 - \theta_2 ),
\cr
\dot{\theta}_2 &= \omega + \epsilon \Gamma( \theta_2 - \theta_1 ),
\end{align}
where $\Gamma(\phi)$ is the phase coupling function defined as
\begin{align}
\Gamma(\phi) 
&= \frac{1}{2\pi} \int_0^{2\pi} {\bm Z}(\phi + \psi) \cdot {\bm H}( \phi +\psi, \psi)  d\psi
\cr
&=
\la {\bm Z}(\phi + \psi) \cdot {\bm H}( \phi +\psi, \psi) \ra_\psi
\cr
&= \la {\bm Z}(\psi) \cdot {\bm H}( \psi, \psi - \phi ) \ra_\psi.
\end{align}
Here, we have defined an average of a function $f(\psi)$ over one period of oscillation as
\begin{align}
\la f(\psi) \ra_{\psi} = \frac{1}{2\pi} \int_0^{2\pi} f(\psi) d\psi.
\end{align}

\subsection{Linear stability of the in-phase synchronized state}

From the coupled phase equations~(\ref{eq7}), the dynamics of the phase difference $\phi = \theta_1 - \theta_2$ obeys
\begin{align}
\label{eq10}
\dot{\phi} = \epsilon [ \Gamma(\phi) - \Gamma(-\phi) ],
\end{align}
where the right-hand side is the antisymmetric part of the phase coupling function $\Gamma(\phi)$.
This equation always has a fixed point at $\phi=0$ corresponding to the in-phase synchronized state. In a small vicinity of $\phi=0$, the above equation can be linearized as
\begin{align}
\dot{\phi} \approx 2 \epsilon \Gamma'(0) \phi.
\end{align}
The derivative of the phase coupling function is given by
\begin{align}
\Gamma'(\phi) =& - \la {\bm Z}(\psi) \cdot {\bm H}'_2(\psi, \psi-\phi) \ra_\psi,
\end{align}
where the partial derivative of ${\bm H}$ with respect to the second argument is denoted as
\begin{align}
{\bm H}_2'(\psi_1, \psi_2) &= \frac{\partial {\bm H}(\psi_1, \psi_2)}{\partial \psi_2}.
\end{align}
Thus, the linear stability of this state is characterized by the exponent $2 \epsilon \Gamma'(0)$, and a larger $-\Gamma'(0)$ yields a higher linear stability of the in-phase synchronized state.

In this study, we consider optimization of the mutual coupling term, ${\bm H}$, or either the parameters or functions included in it, so that the linear stability
\begin{align}
- \Gamma'(0) 
=& \la {\bm Z}(\psi) \cdot {\bm H}_2'(\psi, \psi) \ra_\psi
\end{align}
of the in-phase synchronized state, $\phi = 0$, is maximized under appropriate constraints on the intensity of the mutual coupling.

\begin{figure}[htbp]
\centering
\includegraphics[width=0.8\hsize,clip]{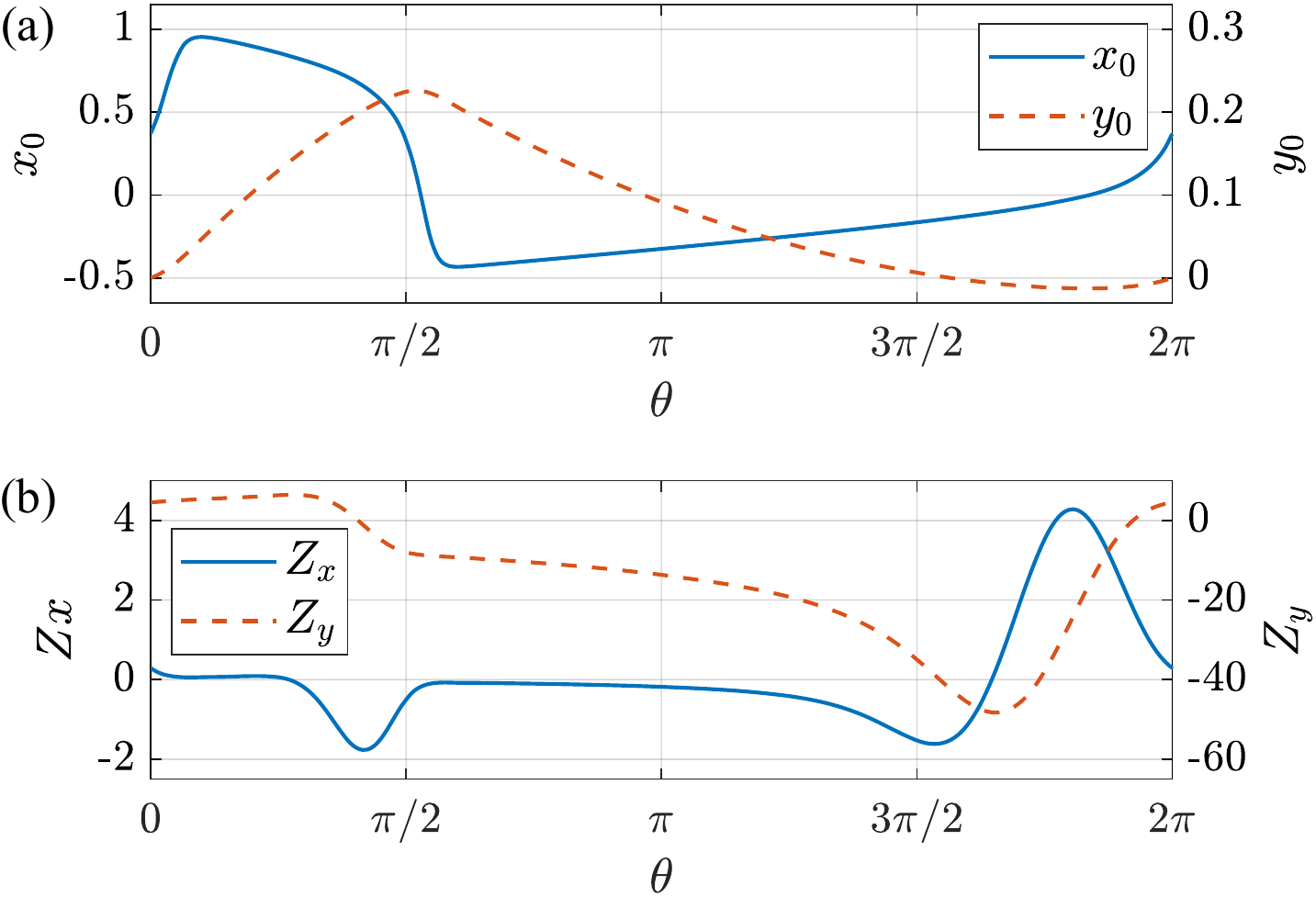}
\caption{Limit-cycle orbit and phase sensitivity function of the FitzHugh-Nagumo oscillator. Time sequences of the $x$ and $y$ components are plotted for one period of oscillation, $0 \leq \theta < 2\pi$. (a) Limit cycle $(x_0(\theta), y_0(\theta))$. (b) Phase sensitivity function $(Z_x(\theta), Z_y(\theta))$.}
\label{fig1}
\end{figure}

\subsection{Examples of limit-cycle oscillators}

The Stuart-Landau (SL) oscillator is used in the following numerical examples, which is a normal form of the supercritical Hopf bifurcation and is
described by
\begin{align}
{\bm X} = \begin{pmatrix} x \\ y \end{pmatrix} \in {\mathbb R}^2,
\end{align}
\begin{align}
{\bm F} = \begin{pmatrix} F_x \\ F_y \end{pmatrix} =
\begin{pmatrix}
{ x - a y - \left( x ^ { 2 } + y ^ { 2 } \right)  ( x - b y )}
\\
{ a x + y - \left( x ^ { 2 } + y ^ { 2 } \right)  ( b x + y )}
\end{pmatrix}
,
\end{align}
where the parameters are fixed at $a=2$ and $b=1$. This oscillator has a stable limit cycle with a natural frequency $\omega = a - b = 1$ and period $T = 2 \pi$, represented by
\begin{align}
{\bm X}_0(\theta) = \begin{pmatrix} x_0(\theta) \\ y_0(\theta) \end{pmatrix} = \begin{pmatrix}  \cos \theta \\ \sin \theta \end{pmatrix}, \quad (0 \leq \theta < 2\pi).
\end{align}
The phase function can be explicitly represented by
\begin{align}
\Theta(x, y) = \arctan \frac{y}{x} - \frac{b}{2} \ln ( x^2 + y^2 )
\end{align}
on the whole $xy$-plane except $(0, 0)$, and the PSF is given by
\begin{align}
{\bm Z}(\theta) = \begin{pmatrix} Z_x \\  Z_y \end{pmatrix} = \left( \begin{array} { c } { - \sin \theta - b \cos \theta } \\ { \cos \theta - b \sin \theta } \end{array} \right).
\end{align}

As another example, we use the FitzHugh-Nagumo (FHN) oscillator, described by
\begin{align}
{\bm X} = \begin{pmatrix} x \\ y \end{pmatrix} \in {\mathbb R}^2,
\end{align}
\begin{align}
{\bm F} = \begin{pmatrix} F_x \\ F_y \end{pmatrix} =
\left( \begin{array} { c } { x ( x - c ) ( 1 - x ) - y }
\\
\mu^{-1} ( x - d y ) 
\end{array}
\right),
\end{align}
where the parameters are fixed at $c = - 0.1$, $d = 0.5$, and $\mu = 100$. As $\mu$ is large, this oscillator is a slow-fast system
whose $x$ variable evolves much faster than the $y$ variable, leading to relaxation oscillations. With these parameters, the natural period of the
oscillation is $T \approx 126.7$ and the natural frequency is $\omega \approx 0.0496$.
The limit cycle ${\bm X}_0(\theta) = (x_0(\theta), y_0(\theta))^{\dag}$ and phase function $\Theta({\bm X})$ cannot be obtained analytically for this model, but the PSF ${\bm Z}(\theta)$ can be obtained by numerically solving the
adjoint equation.
Figure~\ref{fig1} shows the time sequences of the limit-cycle orbit ${\bm X}_0(\theta)$
and PSF ${\bm Z}(\theta)$ for one period of oscillation, $0 \leq \theta < 2\pi$.

\section{Linear coupling}

\subsection{Time-delayed coupling}

First, we consider a simple time-delayed coupling, where each oscillator is driven by the past state of the other oscillator. The model is given by
\begin{align}
\label{eq3}
\dot{\bm X}_1 &= {\bm F}({\bm X}_1) + \epsilon \sqrt{P} {\rm K} {\bm X}_2(t-\tau),
\cr
\dot{\bm X}_2 &= {\bm F}({\bm X}_2) + \epsilon \sqrt{P} {\rm K} {\bm X}_1(t-\tau),
\end{align}
where $0 < \epsilon \ll 1$ is a small parameter representing the strength of the interaction, $P > 0$ is a real constant that controls the norm of the coupling signal, ${\rm K} \in {\mathbb R}^{N \times N}$ is a constant matrix specifying which components of the oscillator states ${\bm X}_{1, 2}(t)$ are coupled, and $\tau$ ($0 \leq \tau \leq T$) is a time delay.
In our previous study~\cite{shirasaka17b}, we considered optimization of the matrix ${\rm K}$ for the case where the two oscillators are nearly identical and coupled without time delay. 
Here, we consider two oscillators with identical properties, fix the matrix ${\rm K}$ constant, and vary the time delay, $\tau$, to improve the linear stability of the in-phase synchronized state.

In this case, the coupling functionals are given by
\begin{align}
\hat{\bm H}\{ {\bm X}_1^{(t)}, {\bm X}_2^{(t)} \} = \sqrt{P} {\rm K} {\bm X}_2(t-\tau),
\cr
\hat{\bm H}\{ {\bm X}_2^{(t)}, {\bm X}_1^{(t)} \} = \sqrt{P} {\rm K} {\bm X}_1(t-\tau),
\end{align}
which can be expressed as functions of $\theta_1$ and $\theta_2$ as
\begin{align}
{\bm H}( \theta_1, \theta_2 ) = \sqrt{P} {\rm K} {\bm X}_0(\theta_2-\omega\tau),
\cr
{\bm H}( \theta_2, \theta_1 ) = \sqrt{P} {\rm K} {\bm X}_0(\theta_1-\omega\tau),
\end{align}
after phase reduction. The phase coupling function is
\begin{align}
\Gamma(\phi)
&= \la {\bm Z}(\psi) \cdot {\bm H}( \psi, \psi - \phi ) \ra_\psi
\cr
&= \la {\bm Z}(\psi) \cdot \sqrt{P} {\rm K} {\bm X}_0(\psi-\phi - \omega \tau) \ra_\psi,
\end{align}
and the linear stability is characterized by
\begin{align}
- \Gamma'(0)
&= \la {\bm Z}(\psi) \cdot {\bm H}_2'(\psi, \psi) \ra_\psi
\cr
&=
\la {\bm Z}(\psi) \cdot \sqrt{P} {\rm K} {\bm X}_0'(\psi-\omega \tau) \ra_\psi,
\label{delay-stability}
\end{align}
where ${\bm X}_0'(\theta) = d{\bm X}_0(\theta) / d\theta$.

The maximum stability is attained only when $\tau$ satisfies
\begin{align}
\frac{\partial}{\partial \tau} \{ - \Gamma'(0) \}
&=
- \omega \la {\bm Z}(\psi) \cdot \sqrt{P} {\rm K} {\bm X}_0''(\psi-\omega \tau) \ra_\psi
= 0,
\label{delay-cond}
\end{align}
where ${\bm X}_0''(\theta) = d^2 {\bm X}_0(\theta)/d\theta^2$. We denote the value of $\tau$ satisfying the above equation as $\tau^*$, i.e.,
\begin{align}
\la {\bm Z}(\psi) \cdot \sqrt{P} {\rm K} {\bm X}_0''(\psi-\omega \tau^*) \ra_\psi = 0.
\end{align}
By partial integration using the $2\pi$-periodicity of ${\bm Z}(\theta)$ and ${\bm X}_0(\theta)$, this can also be expressed as
\begin{align}
\label{eq21}
\la {\bm Z}'(\psi) \cdot \sqrt{P} {\rm K} {\bm X}_0'(\psi-\omega \tau^*) \ra_\psi = 0,
\end{align}
which has the form of a cross-correlation function between ${\bm Z}'(\theta)$ and $\sqrt{P} {\rm K}{\bm X}_0'(\theta)$.
Because both of these functions are $2\pi$-periodic with zero-mean, the left-hand side of Eq.~(\ref{eq21}) is also $2\pi$-periodic with zero mean. Thus, there are at least two values of $\tau$ satisfying the above equation, as long as ${\bm Z}'(\theta)$ and $\sqrt{P} {\rm K}{\bm X}_0'(\theta)$ are non-constant functions (which holds generally for ordinary limit cycles). By choosing an appropriate value of $\tau^*$, the maximum stability is given by
\begin{align}
- \Gamma'(0)
&= \sqrt{P \la {\bm Z}(\psi) \cdot {\rm K} {\bm X}_0'(\psi-\omega \tau^*) \ra_\psi^{2}}.
\end{align}

\subsection{Coupling via linear filtering}

Generalizing the time-delayed coupling, we consider a case in which the past time sequences of both oscillator states are linearly filtered and used as driving signals for the other oscillators. The model is given by
\begin{align}
\dot{\bm X}_1 &= {\bm F}({\bm X}_1) + \epsilon \int_{0}^{T} h(\tau)  {\rm K} {\bm X}_2(t-\tau) d\tau,
\cr
\dot{\bm X}_2 &= {\bm F}({\bm X}_2) + \epsilon \int_{0}^{T} h(\tau) {\rm K} {\bm X}_1(t-\tau) d\tau,
\end{align}
where $h(\tau) : [0, T] \to {\mathbb R}$ is a $T$-periodic real scalar function representing a linear filter, and ${\rm K} \in {\mathbb R}^{N \times N}$ is a constant matrix specifying which components of ${\bm X}$ are coupled. We optimize the linear filter $h(\tau)$ for a given coupling matrix ${\rm K}$ under a constraint specified below.

The coupling functionals are given by
\begin{align}
\hat{\bm H}\{ {\bm X}_1^{(t)}, {\bm X}_2^{(t)} \} = \int_{0}^{T} h(\tau) {\rm K} {\bm X}_2(t-\tau) d\tau,
\cr
\hat{\bm H}\{ {\bm X}_2^{(t)}, {\bm X}_1^{(t)} \} = \int_{0}^{T} h(\tau) {\rm K} {\bm X}_1(t-\tau) d\tau,
\end{align}
which simplify to ordinary functions
\begin{align}
{\bm H}( \theta_1, \theta_2 ) = \int_{0}^{T} h(\tau) {\rm K} {\bm X}_0(\theta_2-\omega \tau) d\tau,
\cr
{\bm H}( \theta_2, \theta_1 ) = \int_{0}^{T} h(\tau) {\rm K} {\bm X}_0(\theta_2-\omega \tau) d\tau,
\end{align}
after phase reduction. The phase coupling function is given by
\begin{align}
\Gamma(\phi)
&= \la {\bm Z}(\phi + \psi) \cdot \int_0^T h(\tau) {\rm K}
{\bm X}_0(\psi - \omega \tau) d\tau \ra_\psi
\cr
&= \la \int_0^T {\bm Z}(\psi) \cdot h(\tau) {\rm K} {\bm X}_0(\psi - \omega \tau - \phi) d\tau \ra_\psi
\label{filter-phasecoupling}
\end{align}
and the linear stability of the in-phase synchronized state is characterized by
\begin{align}
- \Gamma'(0)
&= 
\la
\int_0^T
{\bm Z}(\psi) \cdot h(\tau) {\rm K} {\bm X}_0'(\psi - \omega \tau)
d\tau\ra_\psi.
\end{align}

We constrain the $L^2$-norm $\| h(\tau) \| = \sqrt{ \int_0^T h(\tau)^2 d\tau }$ of the linear filter, $h(\tau)$, as $\| h(\tau) \|^2 = Q$, where $Q > 0$ controls the overall coupling intensity, and seek the optimal $h(\tau)$ that maximizes the linear stability, $-\Gamma'(0)$. That is, we consider an optimization problem:
\begin{align}
\mbox{maximize} \quad - \Gamma'(0) \quad \mbox{subject to} \quad \| h(\tau) \|^2 = Q.
\end{align}
To this end, we define an objective functional as
\begin{align}
S\{ h, \lambda \}
=& - \Gamma'(0) + \lambda ( \| h(\psi) \|^2 - Q )
\cr
=&
\la \int_0^T {\bm Z}(\psi) \cdot h(\tau) {\rm K} {\bm X}_0'(- \omega \tau + \psi) d\tau \ra_\psi 
\cr
&+ \lambda \left( \int_0^{T} h(\tau)^2 d\tau - Q \right),
\end{align}
where $\lambda$ is a Lagrange multiplier. From the extremum condition of $S$, the functional derivative of $S$ with respect to $h(\tau)$ should satisfy
\begin{align}
\frac{\delta S}{\delta h(\tau)} = \la {\bm Z}(\psi) \cdot {\rm K} {\bm X}_0'(- \omega \tau + \psi) \ra_\psi + 2\lambda h(\tau) = 0
\end{align}
and the partial derivative of $S$ by $\lambda$ should satisfy
\begin{align}
\frac{\partial S}{\partial \lambda} = 
\int_0^{T} h(\tau)^2 d\tau - Q = 0.
\end{align}
Thus, the optimal linear filter $h(\tau)$ is given by
\begin{align}
h(\tau)\
= - \frac{1}{2 \lambda} \la {\bm Z}(\psi) \cdot  {\rm K} {\bm X}_0'(\psi - \omega \tau) \ra_{\psi}.
\label{optimalfilter}
\end{align}
The Lagrange multiplier $\lambda$ is determined from the constraint $\| h(\tau) \|^2 = Q$, i.e.,
\begin{align}
\frac{1}{4 \lambda^2} \int_0^T \la 
 {\bm Z}(\psi) \cdot  {\rm K} {\bm X}_0'(\psi - \omega \tau) \ra_{\psi}^2 d\tau  = Q,
\end{align}
as
\begin{align}
\lambda = - \sqrt{ \frac{1}{4 Q} \int_0^T \la {\bm Z}(\psi) \cdot  {\rm K} {\bm X}_0'(\psi - \omega \tau) \ra_{\psi}^2 d\tau  },
\end{align}
where the negative sign should be chosen for the in-phase synchronized state to be linearly stable, $-\Gamma'(0) > 0$. The maximum linear stability with the optimized $h(\tau)$ is
\begin{align}
- \Gamma'(0) 
&=
\sqrt{Q  
	\int_0^T \la {\bm Z}(\psi) \cdot  {\rm K} {\bm X}_0'(\psi - \omega \tau)  \ra_{\psi}^2 d\tau}.
\label{filter-linearstab}
\end{align}
 
\subsection{Numerical examples}

\subsubsection{Time-delayed coupling}

We use the SL and FHN oscillators in the following numerical illustrations. For both models, the coupling matrix is assumed to be
\begin{align}
{\rm K} = 
\begin{pmatrix}
1 & 0\\
0 & 0
\end{pmatrix}.
\label{k-matrix}
\end{align}
We compare the optimized case with the non-optimized case, i.e.,
\begin{align}
\dot{\bm X}_1 &= {\bm F}({\bm X}_1) + \epsilon \sqrt{P} {\rm K} {\bm X}_2(t),
\cr
\dot{\bm X}_2 &= {\bm F}({\bm X}_2) + \epsilon \sqrt{P} {\rm K} {\bm X}_1(t),
\label{non-optimized}
\end{align}
where $\epsilon$ is a small parameter that determines the coupling strength and $P$ control the norm of the coupling signal. The mean square of the coupling term over one-period of oscillation is the same in both cases,
that is,
\begin{align}
\la | \sqrt{P} {\rm K} {\bm X}_0(\psi) |^2 \ra_\psi = \la | \sqrt{P} {\rm K} {\bm X}_0(\psi-\omega\tau^*) |^2 \ra_\psi.
\label{kx-norm}
\end{align}

\begin{figure}[t]
\centering
\includegraphics[width=\hsize,clip]{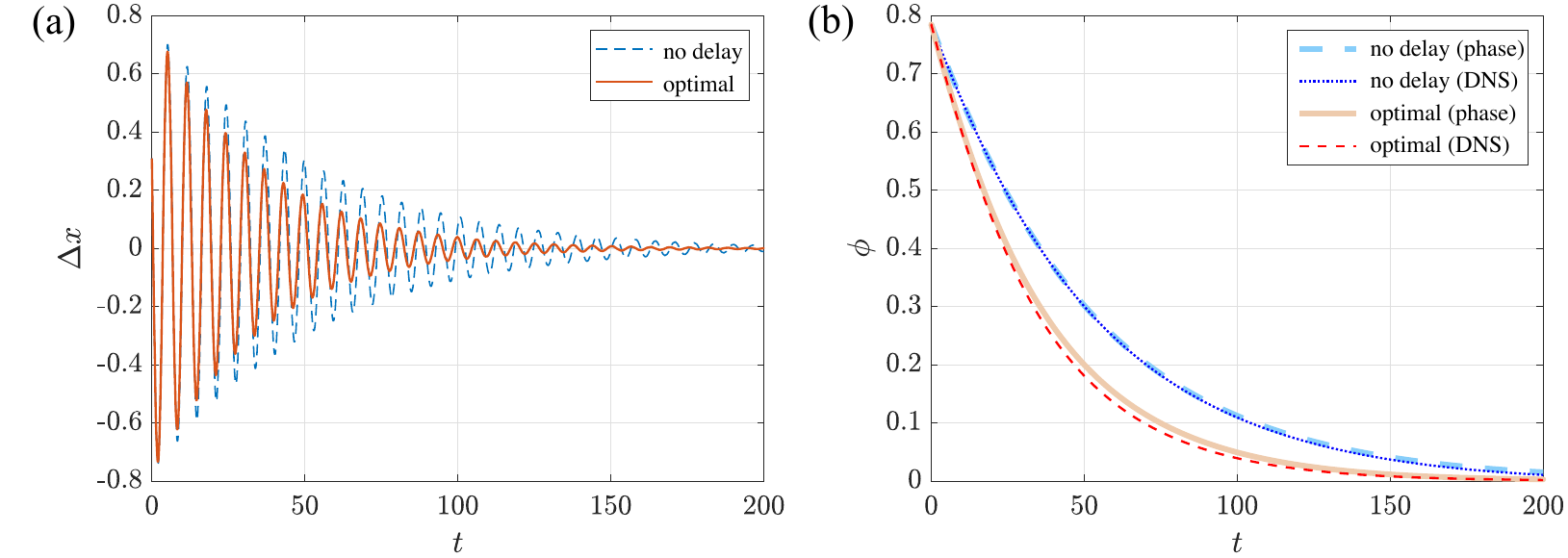}
\caption{Synchronization of two Stuart-Landau oscillators coupled with time delay. The results with the optimal time delay are compared with those without time delay. (a) Evolution of the difference $\Delta x$ between $x$ variables of the two oscillators. (b) Evolution of the phase difference $\phi$ between the oscillators.}
\label{fig2}
\end{figure}
\begin{figure}[htbp]
\centering
\includegraphics[width=\hsize,clip]{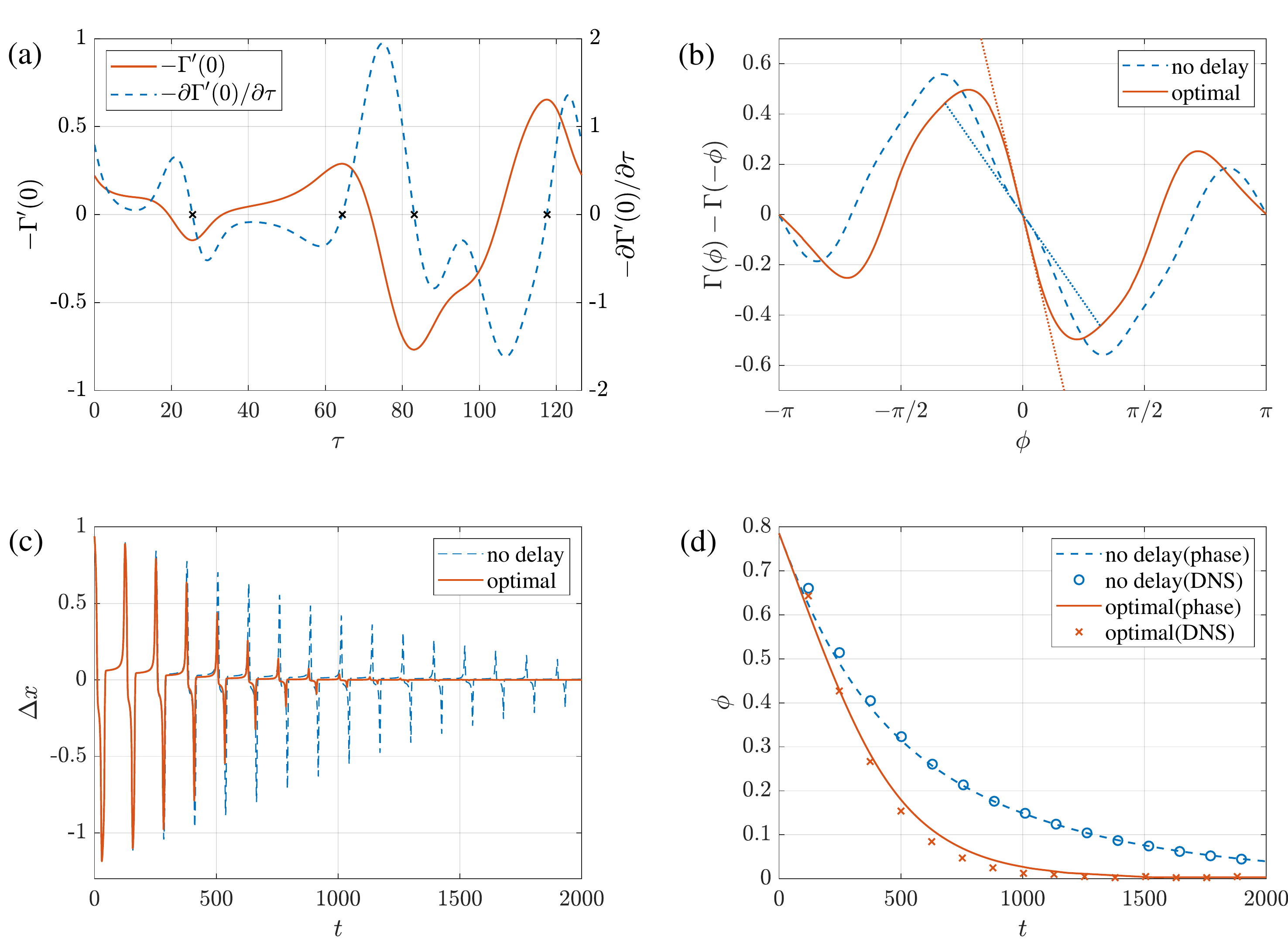}
\caption{Synchronization of two FitzHugh-Nagumo oscillators coupled with time delay. In (b--d), the results with the optimal time delay are compared with those without time delay. (a) Linear stability $- \Gamma'(0)$ and its derivative $-\partial \Gamma'(0) / \partial \tau$ vs. time delay $\tau$. The crosses indicate the values of $\tau$ where $-\partial \Gamma'(0) / \partial \tau = 0$. (b) Antisymmetric part of the phase coupling function, $\Gamma(\phi) - \Gamma(-\phi)$. (c) Evolution of the difference $\Delta x$ between $x$ variables of the two oscillators. (d) Evolution of the phase difference $\phi$ between the oscillators.}
\label{fig3}
\end{figure}

First, for the SL oscillator, we can analytically calculate the optimal time delay. The linear stability of the in-phase synchronized state, Eq.~(\ref{delay-stability}), is given by
\begin{align}
- \Gamma'(0) = 
\sqrt{P} \la Z_x(\psi) x_0'(\psi - \omega \tau) \ra_{\psi}
= \frac{\sqrt{P}}{2} [ \cos (\omega \tau) - b \sin (\omega \tau) ].
\label{sl-delay-stab}
\end{align}
The optimal time delay $\tau = \tau^*$ is determined from Eq.~(\ref{delay-cond}), or equivalently from
\begin{align}
\la Z_x(\psi) x_0''(\psi - \omega \tau) \ra_{\psi} = \frac{b \cos (\omega \tau ) + \sin (\omega \tau)}{2}  = 0.
\end{align}
For the parameter values $b=1$ and $\omega = 1$, this equation is satisfied when $\tau = 3\pi / 4$ or $\tau = 7\pi / 4$. Substituting this into Eq.~(\ref{sl-delay-stab}), we find that $\tau^*=7\pi / 4$ should be chosen, and the maximum linear stability is given by $-\Gamma'(0) = \sqrt{P} / \sqrt{2}$. For the case with no time delay, the linear stability is $-\Gamma'(0) = \sqrt{P} / 2$. Thus, by appropriately choosing the time delay, the linear stability improves by a factor of $\sqrt{2}$ in this case.

Figure~\ref{fig2} shows synchronization of two SL oscillators for the cases with the optimal time delay and without time delay, where $\epsilon = 0.02$, $P=1$, and the initial phase difference is $\phi(0) = \pi / 4$. In Fig.~\ref{fig2}(a), the difference $\Delta x$ between the $x$ variables of the two oscillators, obtained by direct numerical simulations of the coupled SL oscillators, is plotted as a function of $t$. It can be seen that the in-phase synchronized state is established faster in the optimized case because of the higher linear stability. Figure~\ref{fig2}(b) shows the convergence of the phase difference $\phi$ to $0$. It can be seen that the results of the reduced phase equation agree well with those of direct numerical simulations.

Figure~\ref{fig3} shows the results for two FHN oscillators, where $\epsilon = 0.003$, $P=1$, and the initial phase difference is $\phi = \pi / 4$. Figure~\ref{fig3}(a) plots the linear stability $-\Gamma'(0)$ and its derivative $-\partial \Gamma'(0) / \partial \tau$ as functions of the time delay $\tau$, where there are two extrema of $-\Gamma'(0)$. We choose the larger extremum, which is attained at the optimal time delay $\tau^* \approx 117.6$. The antisymmetric part of the phase coupling function, $\Gamma(\phi) - \Gamma(-\phi)$, is shown in Fig~\ref{fig3}(b) for the cases with the optimal delay and without delay. It can be seen that the stability of the in-phase synchronized state $\phi=0$ is improved, as indicated by the straight lines in Fig.~\ref{fig3}(b), where $-\Gamma'(0)\approx0.654$ with the optimized time delay and $-\Gamma'(0)\approx0.221$ without the time delay. The evolution of the difference $\Delta x$ between the $x$ variables of the oscillators is plotted as a function of $t$ in Fig.~\ref{fig3}(c). The phase differences $\phi$ converging toward $0$, obtained from the phase equation and direct numerical simulations of the original model, are shown in Fig.~\ref{fig3}(d). It can be seen that the convergence to the in-phase synchronization is faster with the optimized time delay, and the results of the reduced phase equation agree well with direct numerical simulations.

\subsubsection{Coupling via linear filtering}

We also assume that the coupling matrix ${\rm K}$ is given by Eq.~(\ref{k-matrix}), and compare the results for the optimized case with linear filtering with those for the non-filtered case given by Eq.~(\ref{non-optimized}).
We choose the parameter $Q$ that constrains the norm of the linear filter so that the squared average of the coupling term over one-period of oscillation becomes equal to that in the non-filtered case given by Eq.~(\ref{non-optimized}), i.e.,
\begin{align}
\la \left| \int_0^T h(\tau) {\rm K} {\bm X}_0(\psi-\omega\tau) d\tau \right|^2 \ra_\psi = \la | \sqrt{P} {\rm K} {\bm X}_0(\psi) |^2 \ra_\psi.
\label{kxh-norm}
\end{align}

For the SL oscillators, the optimal filter $h(\tau)$, Eq.~(\ref{optimalfilter}), is explicitly calculated as
\begin{align}
h(\tau) = \sqrt{\frac{Q \omega }{\pi (1 +  b^2 ) }} [ \cos ( \omega \tau  ) - b \sin ( \omega \tau ) ].
\end{align}
The optimal phase coupling function, Eq.~(\ref{filter-phasecoupling}), and optimized linear stability, Eq.~(\ref{filter-linearstab}), are expressed as
\begin{align}
\Gamma(\phi) = -\frac{1}{2} \sqrt{\frac{\pi \left(1+b^2\right) Q}{\omega }}  \sin \phi,
\end{align}
and
\begin{align}
-\Gamma'(0) = \frac{1}{2} \sqrt{\frac{\pi \left(1+b^2\right) Q}{\omega }},
\end{align}
respectively. We take $Q = \omega P / \pi$ so that Eq.~(\ref{kxh-norm}) is satisfied.
The linear stability is then $-\Gamma'(0) = \sqrt{\left(1+b^2\right) P} / 2$ when the optimized linear filter is used and $-\Gamma'(0) = \sqrt{P}/2$ when no filtering of the oscillator state is performed.
Thus, the linear stability is improved by a factor of $\sqrt{2}$ when $b=1$.

It is important to note that, in the SL oscillator case, ${\bm X}_0(\psi)$, ${\bm Z}(\psi)$,
and hence the linear filter $h(\tau)$ contain only the fundamental frequency, i.e., they are purely sinusoidal. Thus, the linear filtering can only shift the phase of the coupling signal and gives the same result as the previous case with the simple time delay. It is also interesting to note that the stability cannot be improved (it is already optimal without filtering) when the parameter $b$, which characterizes non-isochronicity of the limit cycle, is zero.

\begin{figure}[htbp]
\centering
\includegraphics[width=\hsize,clip]{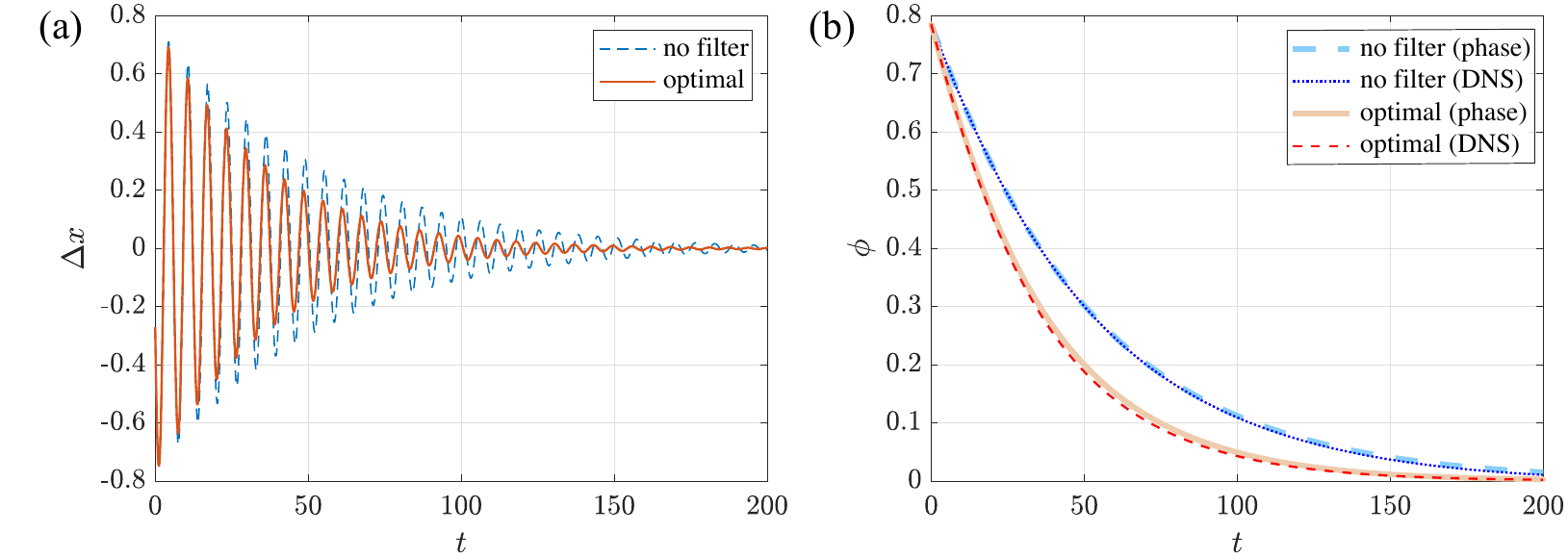}
\caption{Synchronization of two Stuart-Landau oscillators coupled with linear filtering. The results with the optimal filtering are compared with those without filtering. (a) Evolution of the difference $\Delta x$ in the $x$ variables between the oscillators. (b) Evolution of the phase difference $\phi$ between the oscillators.}
\label{fig4}
\end{figure}

Figure~\ref{fig4} shows the synchronization of two SL oscillators, with and without linear filtering, where $\epsilon = 0.02$, $P=1$, and the initial phase difference is $\phi = \pi / 4$. Figure~\ref{fig4}(a) shows evolution of the difference $\Delta x$ between the $x$ variables of the oscillators, and Fig.~\ref{fig4}(b) shows the convergence of the phase difference $\phi$ to $0$. We can see that the in-phase synchronized state is established faster in the optimized case, and the results of the reduced phase model and direct numerical simulations agree well.

\begin{figure}[htbp]
\centering
\includegraphics[width=\hsize,clip]{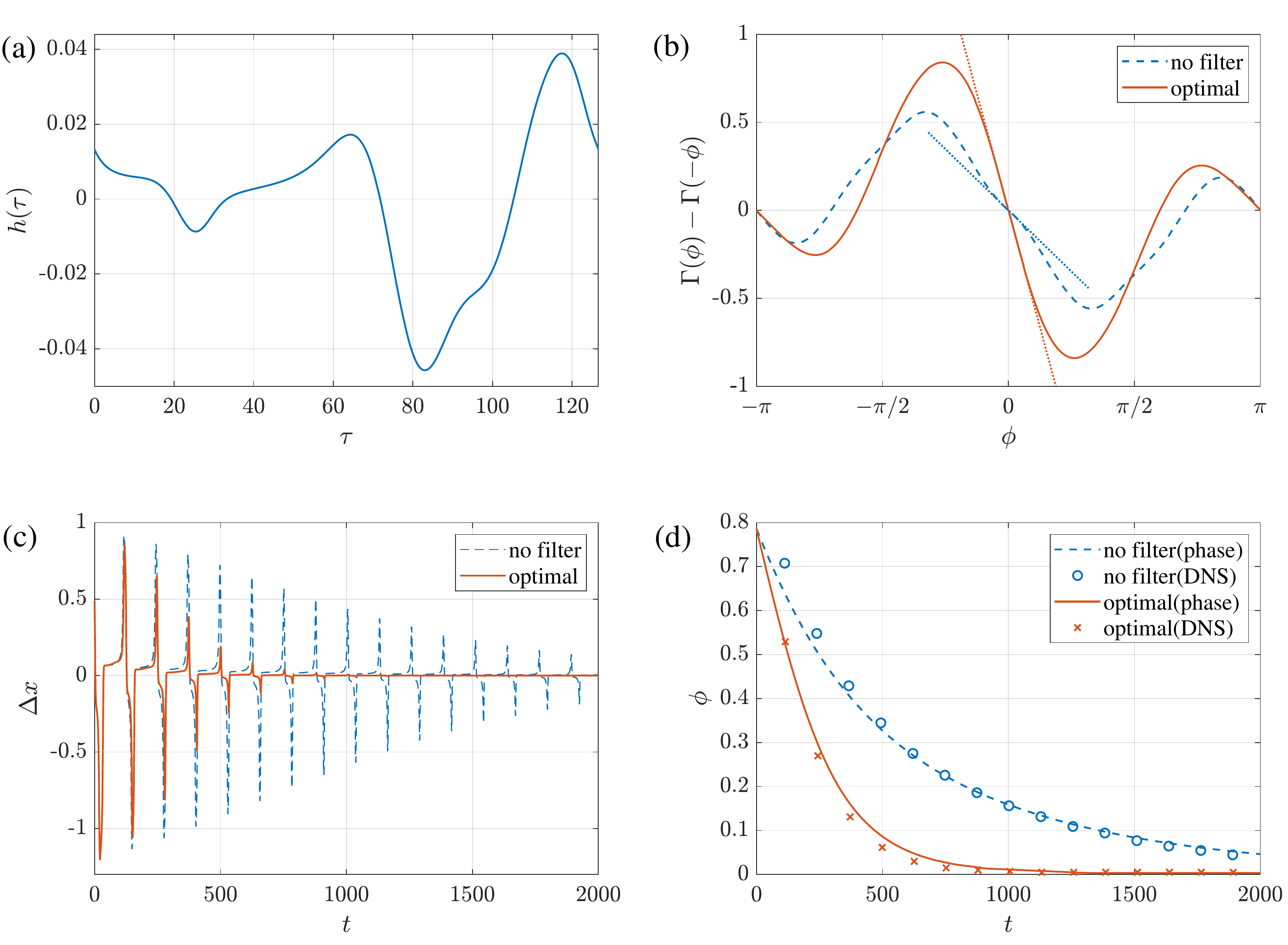}
\caption{Synchronization of two FitzHugh-Nagumo oscillators coupled with linear filtering. In (b--d), the results with the optimal filtering are compared with those without filtering. (a) Optimal linear filter $h(\tau)$. (b) Antisymmetric part of the phase coupling function $\Gamma(\phi) - \Gamma(-\phi)$. (c) Evolution of the difference $\Delta x$ in the $x$ variables between the oscillators. (d) Evolution of the phase difference $\phi$ between the oscillators.}
\label{fig5}
\end{figure}

For the FHN oscillators, the optimal linear filter can be calculated from the time sequences of the limit-cycle solution and PSF obtained numerically. Figure~\ref{fig5} shows the synchronization of two coupled FHN oscillators, with and without linear filtering, where
$\epsilon = 0.003$, $P=1$, $Q \approx 0.0522$, and the initial phase difference is $\phi = \pi / 4$.
Figure~\ref{fig5}(a) shows the optimal filter, (b) antisymmetric part $\Gamma(\phi) - \Gamma(-\phi)$ of the phase coupling function $\Gamma(\phi)$, (c) evolution of the difference $\Delta x$ between the $x$ variables of the oscillators, and (d) convergence of the phase difference $\phi$ toward $0$.
The linear stability is given by 
$-\Gamma'(0)\approx0.844$
for the case with the optimal filter and by
$-\Gamma'(0)\approx0.221$
for the case without filtering, as shown by the straight lines in Fig.~\ref{fig5}(b).
The in-phase synchronized state is established faster in the optimized case, and the results of the reduced phase model and direct numerical simulations agree well.
Because the FHN oscillator has the higher harmonic components in ${\bm X}_0(\psi)$ and ${\bm Z}(\psi)$, the optimal filter $h(\tau)$ can exploit these components, and hence the improvement in the linear stability is larger than that for the case with simple delay.


\section{Nonlinear coupling}

\subsection{Mutual drive-response coupling}

In this section, we consider the case of oscillators interacting through nonlinear coupling functionals.
We assume that the coupling is of a drive-response type, i.e., it can be written as a product of a response matrix of the driven oscillator and a driving function that transforms the signal from the other oscillator.
The model is given by
\begin{align}
\label{eq44}
\dot{\bm X}_1(t) &= {\bm F}({\bm X}_1(t)) + \epsilon \hat{\rm A}\{ {\bm X}_1^{(t)}(\cdot) \} \hat{\bm G} \{ {\bm X}_2^{(t)}(\cdot) \},
\cr
\dot{\bm X}_2(t) &= {\bm F}({\bm X}_2(t)) + \epsilon \hat{\rm A}\{ {\bm X}_2^{(t)}(\cdot) \} \hat{\bm G} \{ {\bm X}_1^{(t)}(\cdot) \},
\end{align}
where the matrix $\hat{\rm A} : C \to {\mathbb R}^{N \times N}$ is a functional of the time sequence of each oscillator representing its response properties and $\hat{\bm G} : C \to {\mathbb R}^N$ is a functional that transforms the time sequence of the other oscillator to a driving signal. 
It should be noted that we may also include self-coupling terms of the form $\epsilon \hat{\bm I}\{{\bm X}^{(t)}(\cdot)\}_{1,2}$ to each equation, which allows the analysis, for example, of diffusive coupling that depends on the state difference between the oscillators. As explained in the Appendix A, it can be shown that inclusion of such self-coupling terms does not alter the results and the linear stability remains the same in the phase-reduction approximation. We thus analyze Eq.~(\ref{eq44}) hereafter.

The coupling functionals in this case are given by
\begin{align}
\hat{\bm H}\{ {\bm X}_1^{(t)}, {\bm X}_2^{(t)} \} = \hat{\rm A}\{ {\bm X}_1^{(t)}(\cdot) \} \hat{\bm G} \{ {\bm X}_2^{(t)}(\cdot) \},
\cr
\hat{\bm H}\{ {\bm X}_2^{(t)}, {\bm X}_1^{(t)} \} = \hat{\rm A}\{ {\bm X}_2^{(t)}(\cdot) \} \hat{\bm G} \{ {\bm X}_1^{(t)}(\cdot) \},
\end{align}
and, as argued in Sec.~IIB, at the lowest-order phase reduction, these functionals can be expressed as ordinary functions of the phase $\theta_1$ and $\theta_2$ as
\begin{align}
{\bm H}( \theta_1, \theta_2 ) = {\rm A}( \theta_1 ) {\bm G} ( \theta_2 ),
\cr
{\bm H}( \theta_2, \theta_1 ) = {\rm A}( \theta_2 ) {\bm G} ( \theta_1 ),
\end{align}
where we introduced ordinary $2\pi$-periodic functions ${\rm A}$ and ${\bm G}$ of $\theta_1$ and $\theta_2$.
Using these functions, the phase coupling function is given by
\begin{align}
\Gamma(\phi)
&= \la {\bm Z}(\psi) \cdot {\rm A}(\psi){\bm G}(\psi-\phi) \ra_\psi,
\end{align}
and the linear stability is characterized by
\begin{align}
-\Gamma'(0) 
=& \la {\bm Z}(\psi) \cdot {\rm A}(\psi){\bm G}'(\psi) \ra_\psi
\cr
=& \la {\rm A}^{\dag}(\psi) {\bm Z}(\psi) \cdot {\bm G}'(\psi) \ra_\psi
\cr
=& - \la \frac{d}{d\psi} \left[ {\rm A}^{\dag}(\psi) {\bm Z}(\psi) \right] \cdot {\bm G}(\psi) \ra_\psi,
\end{align}
where the last expression is obtained by partial integration using $2\pi$-periodicity of ${\rm A}(\psi)$, ${\bm Z}(\psi)$, and ${\bm G}(\psi)$.

Therefore, although we started from Eq.~(\ref{eq44}) with a general drive-response coupling that depends on the past time sequences of the oscillators, the linear stability can be represented only by the present phase values of the oscillators at the lowest-order phase reduction. 
In the following subsections, we consider the optimization of the response matrix ${\rm A}(\psi)$ or the driving function ${\bm G}(\psi)$, represented as functions of the phase $\psi$.

\subsection{Optimal response matrix}

As for the first case, we optimize the response matrix ${\rm A}(\psi)$ as a function of the phase $\psi$, assuming that the driving functional $\hat{\bm G}$ is given. We introduce a constraint that the squared Frobenius norm of ${\rm A}(\psi)$ averaged over one period of oscillation is fixed as $\la \| {\rm A}(\psi) \|^2 \ra_\psi = P$, and consider an optimization problem:
\begin{align}
\mbox{maximize} \quad -\Gamma'(0) \quad \mbox{subject to} \quad \la \| {\rm A}(\psi) \|^2 \ra_\psi = P,
\end{align}
where $ \| {\rm A} \| = \sqrt{ \sum_{i,j} A_{ij}^2 }$ represents the Frobenius norm of the matrix ${\rm A} = ( A_{ij} )$.
By defining an objective functional,
\begin{align}
S\{ {\rm A}, \lambda \}
&= - \Gamma'(0) + \lambda \left( \la \| {\rm A}(\psi) \|^2 \ra_\psi - P \right)
\cr
&= \la {\bm Z}(\psi) \cdot {\rm A}(\psi) {\bm G}'(\psi) \ra_\psi + \lambda \left( \la \| {\rm A}(\psi) \|^2 \ra_\psi - P \right),
\cr
\end{align}
where $\lambda$ is a Lagrange multiplier, and by taking the functional derivative with respect to each component, $A_{ij}$, of ${\rm A}$, we obtain the extremum condition. In this case, 
\begin{align}
\frac{\delta S}{\delta A_{ij}(\psi)} = \frac{1}{2\pi} Z_i(\psi) G_j'(\psi) + \frac{\lambda}{\pi} A_{ij}(\psi) = 0,
\end{align}
and we obtain
\begin{align}
A_{ij}(\psi) = - \frac{1}{2\lambda} Z_i(\psi) G_j'(\psi)
\end{align}
i.e.,
\begin{align}
{\rm A}(\psi) = - \frac{1}{2\lambda} {\bm Z}(\psi) {\bm G}'(\psi)^{\dag},
\end{align}
and the Lagrange multiplier is determined from the constraint,
\begin{align}
\la \| {\rm A}(\psi) \|^2 \ra_{\psi}
= \frac{1}{4\lambda^2} \la \| {\bm Z}(\psi) {\bm G}'(\psi)^{\dag} \|^2 \ra_\psi
= P
\end{align}
as
\begin{align}
\lambda
&= - \sqrt{ \frac{1}{4P} \la \| {\bm Z}(\psi) {\bm G}'(\psi)^{\dag} \|^2 \ra_\psi },
\end{align}
where the negative sign is chosen so that $-\Gamma'(0) > 0$. The maximum stability of the in-phase synchronized state is
\begin{align}
- \Gamma'(0)
&= \sqrt{ P \la \| {\bm Z}(\psi) {\bm G}'(\psi)^{\dag} \|^2 \ra_\psi }.
\end{align}

\subsection{Optimal driving functional}

We can also seek the function ${\bm G}(\psi)$ that provides the optimal driving signal as a function of the phase $\psi$, assuming that the response matrix $\hat{\rm A}$ is given.
We constrain the squared average of ${\bm G}(\psi)$ over one period of oscillation as $\la | {\bm G}(\psi) |^2 \ra_\psi = P$, and maximize the linear stability of the in-phase state:
\begin{align}
\mbox{maximize} \quad -\Gamma'(0) \quad \mbox{subject to} \quad \la | {\bm G}(\psi) |^2 \ra_\psi = P.
\end{align}

We define an objective functional,
\begin{align}
S\{ {\rm A}, \lambda \}
=& - \Gamma'(0) + \lambda \left( \la | {\bm G}(\psi) |^2 \ra_\psi - P \right)
\cr
=& - \la \frac{d}{d\psi} \left[ {\rm A}^{\dag}(\psi) {\bm Z}(\psi) \right] \cdot {\bm G}(\psi) \ra_\psi 
\cr
&+ \lambda \left( \la | {\bm G}(\psi) |^2 \ra_\psi - P \right),
\cr
\end{align}
where $\lambda$ is a Lagrange multiplier. From the extremum condition for $S$, we obtain
\begin{align}
\frac{\delta S}{\delta {\bm G}(\psi)} = - \frac{1}{2\pi} \frac{d}{d\psi} [ {\rm A}^{\dag}(\psi) {\bm Z}(\psi) ] + \frac{\lambda}{\pi} {\bm G}(\psi) = 0
\end{align}
and the constraint on ${\bm G}$. The optimal driving function is given by
\begin{align}
{\bm G}(\psi) = \frac{1}{2\lambda} \frac{d}{d\psi} [ {\rm A}^{\dag}(\psi) {\bm Z}(\psi) ],
\label{opt-driv}
\end{align}
where the Lagrange multiplier $\lambda$ should be chosen to satisfy the norm constraint,
\begin{align}
\frac{1}{4 \lambda^2} \la \left| \frac{d}{d\psi} [ {\rm A}^{\dag}(\psi) {\bm Z}(\psi) ] \right|^2 \ra_\psi = P.
\end{align}
This yields
\begin{align}
\lambda = - \sqrt{ \frac{1}{4 P} \la \left| \frac{d}{d\psi} [ {\rm A}^{\dag}(\psi) {\bm Z}(\psi) ] \right|^2 \ra_\psi },
\end{align}
where the negative sign is taken to satisfy $\Gamma'(0)<0$. The maximum stability is
\begin{align}
- \Gamma'(0)
&=
\sqrt{ P \la \left| \frac{d}{d\psi} [ {\rm A}^{\dag}(\psi) {\bm Z}(\psi) ] \right|^2 \ra_\psi }.
\end{align}

\subsection{Numerical examples}

\subsubsection{Optimal response matrix}

As an example, we assume that the driving functional $\hat{\bm G} \{ {\bm X}^{(t)}(\cdot) \}$ is simply given by $\hat{\bm G} \{ {\bm X}^{(t)}(\cdot) \} = {\bm X}(t)$, and seek the optimal response matrix ${\rm A}(\psi)$ satisfying $\la \| {\rm A}(\psi) \|^2 \ra_\psi = P$. For comparison, we also consider an identity response matrix, ${\rm A}_I = \mbox{diag}(\sqrt{P/2}, \sqrt{P/2})$, normalized to satisfy $\la \| {\rm A}_I \|^2 \ra_\psi = P$. Note that both the $x$ and $y$ components are coupled, in contrast to the previous section where only the $x$ component is coupled.

For the SL oscillator, the optimal response matrix can be analytically expressed as
\begin{align}
&
{\rm A}(\psi) = \sqrt{\frac{P}{1+b^2}}
\cr
&\times
\left(
\begin{array}{rr}
 \sin \psi  (b \cos \psi +\sin \psi ) & -\cos \psi  (b \cos \psi +\sin \psi ) \\
 \sin \psi  (b \sin \psi -\cos \psi ) & \cos \psi  (\cos \psi -b \sin \psi  )
\end{array}
\right),
\cr
\end{align}
and the phase coupling function is given by $\Gamma(\phi) = - \sqrt{ (1 + b^2) P } \sin \phi$,
which gives the optimal linear stability $-\Gamma'(0) = \sqrt{ ( 1 + b^2 ) P }$.
In contrast, for the identity matrix ${\rm A}_I$, the phase coupling function is
$\Gamma(\phi) = -\sqrt{P/2} (b \cos \phi +\sin \phi )$
and the linear stability is $-\Gamma'(0) = \sqrt{P}/\sqrt{2}$.
Thus, the linear stability is improved by a factor of $\sqrt{ 2 (1+b^2) }$.

\begin{figure}[htbp]
\centering
\includegraphics[width=\hsize,clip]{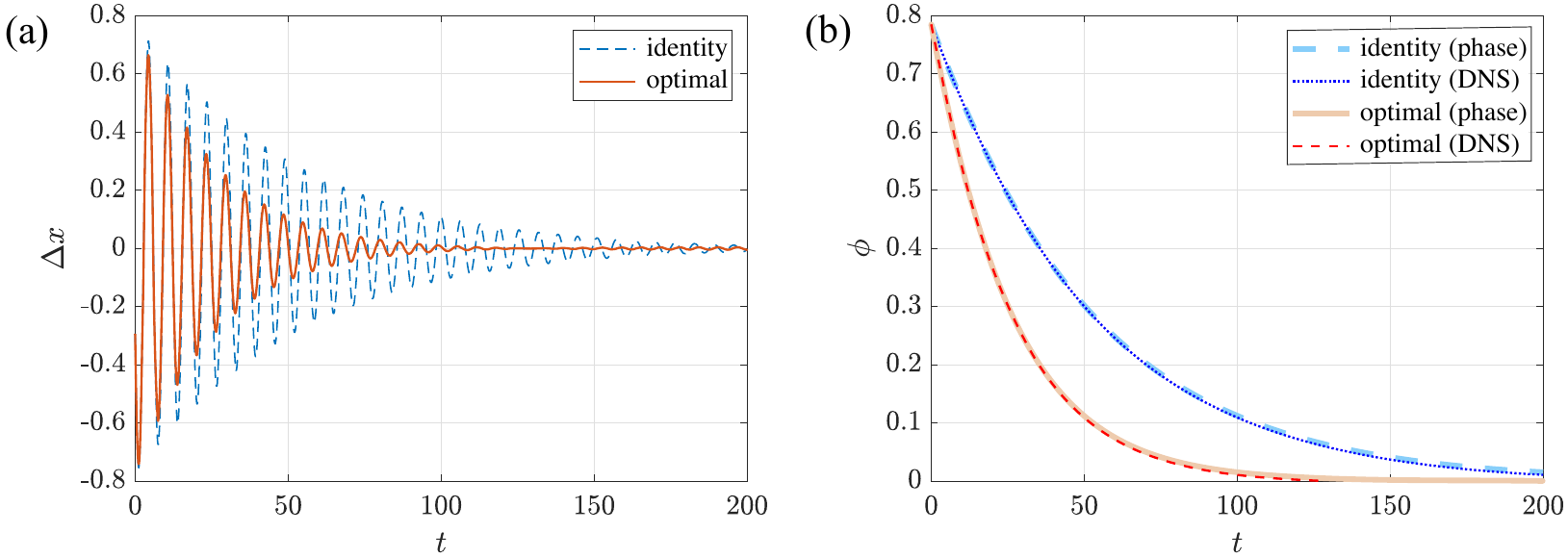}
\caption{Synchronization of Stuart-Landau oscillators with the optimal response matrix. (a) Evolution of the difference $\Delta x$ between the $x$ variables of the oscillators. (b) Evolution of the phase difference $\phi$ between the oscillators.}
\label{fig6}
\end{figure}
\begin{figure}[htbp]
\centering
\includegraphics[width=\hsize,clip]{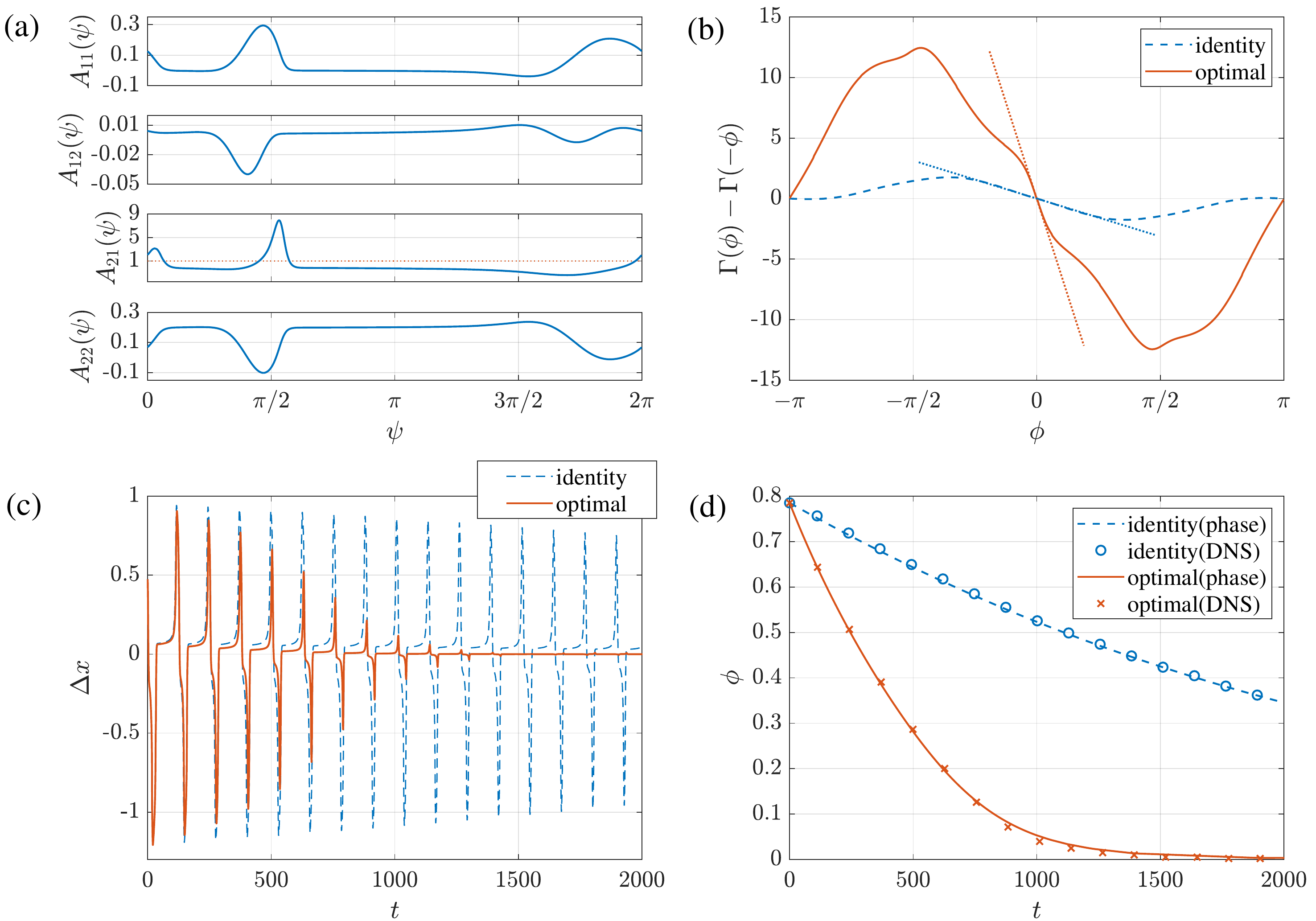}
\caption{Synchronization of FitzHugh-Nagumo oscillators with the optimal response matrix. In (b--d), the results with the optimal response matrix are compared with those with the identity response matrix. (a) Four components of the optimal response matrix ${\rm A}(\psi)$. (b) Antisymmetric part of the phase coupling function, $\Gamma(\phi) - \Gamma(-\phi)$. (c) Evolution of the difference $\Delta x$ between the $x$ variables of the oscillators. (d) Evolution of the phase difference $\phi$ between the oscillators.}
\label{fig7}
\end{figure}

Figure~\ref{fig6} shows synchronization of two SL oscillators for the cases with the optimal response matrix ${\rm A}(\psi)$ and with the identity matrix ${\rm A}_I(\psi)$, where $b=1$, $\epsilon = 0.01$, $P=2$,
and the initial phase difference is $\phi = \pi / 4$. Figure~\ref{fig6}(a) shows the evolution of the difference $\Delta x$ in the $x$ variables between the two oscillators, and Fig.~\ref{fig6}(b) shows the convergence of the phase difference $\phi$ to $0$. The in-phase synchronized state is more quickly established in the optimized case, and the results of the reduced phase model and direct numerical simulations agree well.

For the FHN oscillator, the optimal response matrix can be calculated numerically. Figure~\ref{fig7} compares the synchronization dynamics of two coupled FHN oscillators with the optimal and identity response matrices, where $\epsilon = 0.0002$, $P=2$, and the initial phase difference is $\phi = \pi / 4$. Figure~\ref{fig7}(a) shows four components of the optimal response matrix ${\rm A}(\psi)$ for $0 \leq \psi < 2\pi$.
It is notable that the magnitude of $A_{21}(\psi)$ is much larger than the other components, indicating that driving the $y$ component of each oscillator by using the $x$ component of the other oscillator is efficient in synchronizing the oscillators in this case.
Figure~\ref{fig7}(b) plots the antisymmetric part of the phase coupling functions for the optimal and identity response matrices, which shows that a much higher stability is attained in the optimized case ($-\Gamma'(0)\approx10.1$ for the optimized response matrix and $-\Gamma'(0)\approx0.999$ for the identity response matrix).

Figure~\ref{fig7}(c) shows the time evolution of the difference $\Delta x$ between the two oscillators, and Fig.~\ref{fig7}(d) shows the convergence of the phase difference $\phi$ to $0$.
In order to use the optimal response matrix, instantaneous phase values of the oscillators are necessary. In the direct numerical simulations shown here, we approximately evaluated the phase value by linearly interpolating two consecutive crossings times of the oscillator state at an appropriate Poincar\'e section, and this value was used to generate the driving signal.
It can be seen from the figures that the in-phase synchronized state is established much faster in the optimized case, and the results of the reduced phase model and direct numerical simulations agree well.

\subsubsection{Optimal driving functional}

\begin{figure}[t]
\centering
\includegraphics[width=\hsize,clip]{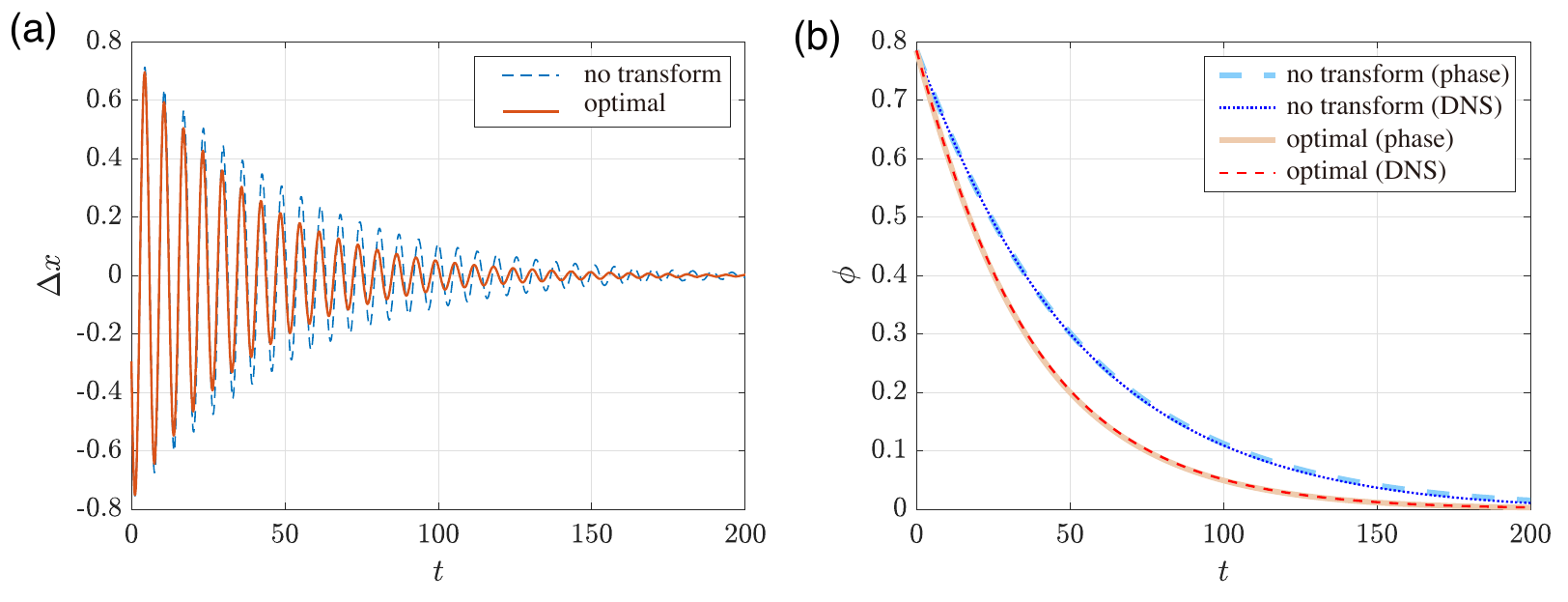}
\caption{Synchronization of Stuart-Landau oscillators coupled with the optimal driving function. (a) Evolution of the difference $\Delta x$ between the $x$ variables of the oscillators. (b) Evolution of the phase difference $\phi$ between the oscillators.}
\label{fig8}
\end{figure}

For the numerical simulations, we assume that ${\rm A}(\psi)$ is simply given by an identity matrix, $\mbox{diag}(1, 1)$.
The optimal driving function ${\bm G}(\psi)$ is then simply given as ${\bm G}(\psi) \propto {\bm Z}'(\psi)$ from Eq.~(\ref{opt-driv}), with the norm constraint $\la | {\bm G}(\psi) |^2 \ra_{\psi} = P$. 
For the SL oscillator, the optimal driving function is explicitly given by
\begin{align}
{\bm G}(\psi) = 
\sqrt{\frac{P}{1+b^2}} 
\left(
\begin{array}{c}
\cos \psi - b \sin \psi \\
b \cos \psi + \sin \psi \\
\end{array}
\right).
\end{align}
Figure~\ref{fig8} shows synchronization of two SL oscillators coupled through the optimal driving function, and coupled without transformation of the oscillator state, i.e., $\hat{\bm G} \{ {\bm X}^{(t)}(\cdot) \} = {\bm X}(t)$, where $b=1$, $\epsilon  = 0.01$, and $P=1$, and the initial phase difference is $\phi = \pi / 4$.
Figure~\ref{fig8}(a) shows the evolution of the difference $\Delta x$ between the $x$ variables of the two oscillators, and Fig.~\ref{fig8}(b) shows the convergence of the phase difference $\phi$ to $0$. It is confirmed that the linear stability of the in-phase synchronized state is improved in the optimized case, and the  results of the reduced phase model and direct numerical simulations agree well.

\begin{figure}[t]
\centering
\includegraphics[width=\hsize,clip]{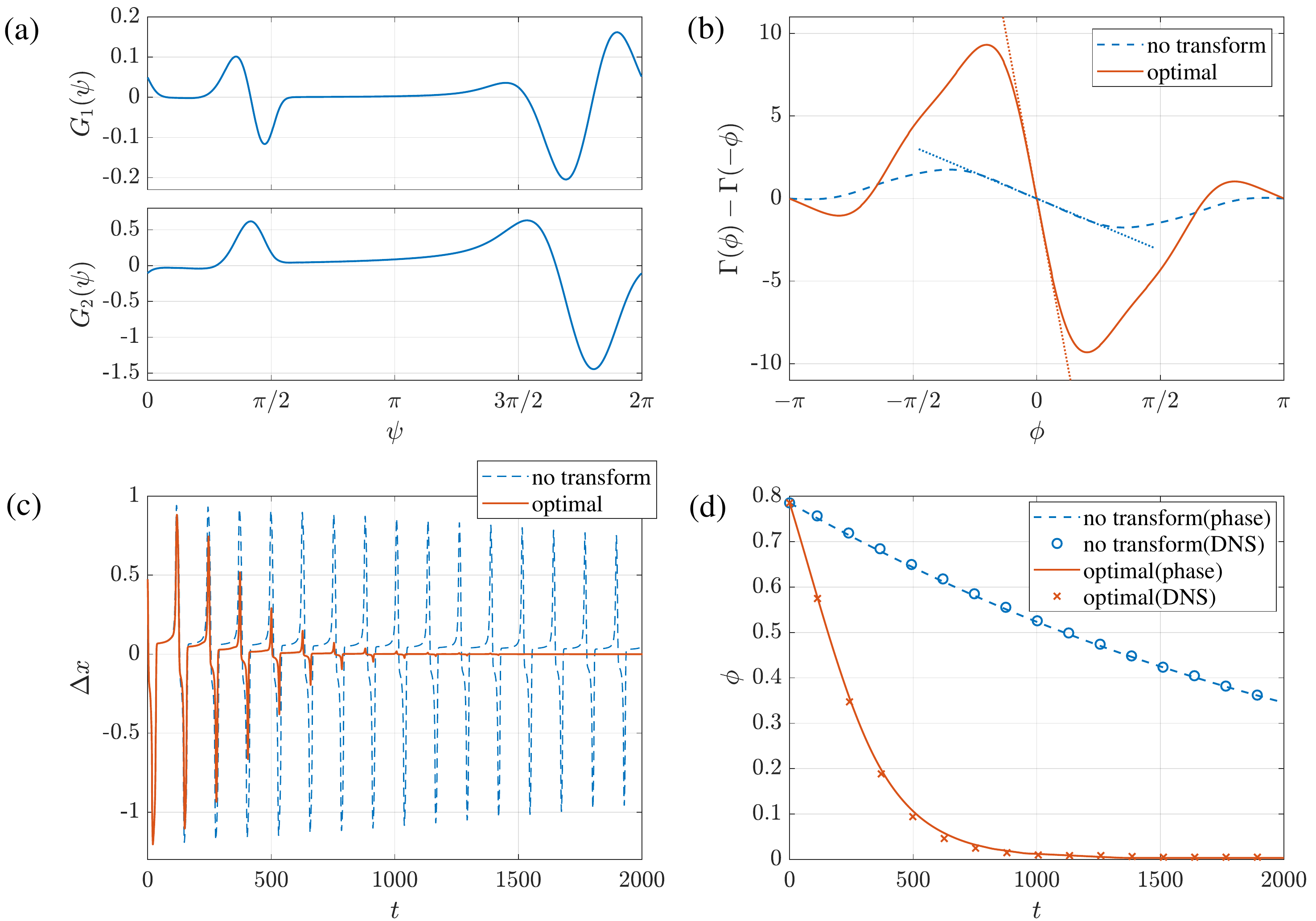}
\caption{Synchronization of FitzHugh-Nagumo oscillators coupled with the optimal driving function. In (b--d), the results with the optimal driving function are compared with those without transformation. (a) Optimal driving function ${\bm G}(\psi) = (G_1(\psi), G_2(\psi))$. (b) Antisymmetric part of the phase coupling function, $\Gamma(\phi) - \Gamma(-\phi)$. (c) Evolution of the difference $\Delta x$ between the $x$ variables of the oscillators. (d) Evolution of the phase difference $\phi$.}
\label{fig9}
\end{figure}

For the FHN oscillator, the norm of ${\bm X}_0(\psi)$ is $\la | {\bm X}_0(\psi) |^2 \ra_{\psi} \approx 0.221$, and we fix the norm $P$ of ${\bm G}(\psi)$ to this value. The optimal driving function can be calculated from ${\bm X}_0(\psi)$ and ${\bm Z}(\psi)$ obtained numerically.
Figure~\ref{fig9} shows synchronization of two FHN oscillators coupled with the optimal driving function, as well as comparison with the non-transformed case, where $\epsilon = 0.0002$, $P \approx 0.221$, and the initial phase difference is $\phi = \pi / 4$. 
Figure~\ref{fig9}(a) shows the optimal driving function ${\bm G}(\psi)$ for $0 \leq \psi < 2\pi$, which is proportional to the derivative ${\bm Z}'(\psi)$. Figure~\ref{fig9}(b) plots the antisymmetric part of the phase coupling function for the optimal driving function ${\bm G}(\psi)$ with that without transformation, respectively, indicating a much higher linear stability in the optimized case
($-\Gamma'(0)\approx12.8$ with the optimized driving function and $-\Gamma'(0)\approx0.999$ without optimization).

Figure~\ref{fig9}(c) shows a plot of the evolution of the difference $\Delta x$ between $x$ variables of the oscillators, and Fig.~\ref{fig9}(d) shows the convergence of the phase difference $\phi$ to $0$.
Similar to the previous case with the optimal response matrix, instantaneous phase values of the oscillators are approximately evaluated by linear interpolation and used to generate the optimal driving signal in the direct numerical simulations.
We can confirm that the in-phase synchronized state is established much faster in the optimized case, and the results of the reduced phase model and direct numerical simulations agree well.


\section{Discussion}

We have shown that, by optimizing the mutual coupling between coupled oscillators, the linear stability of the in-phase synchronized state can be improved and faster convergence to the synchronization can be achieved.
We have shown that, even if we start from a system of coupled oscillators with general coupling functionals that depend on the past time sequences of the oscillators, the system can be approximately reduced to a pair of simple ordinary differential equations that depend only on the present phase values of the oscillators within the phase reduction theory, and the optimal coupling function between the oscillators can be obtained as a function of the phase values.
Though  we have considered only the simplest cases where two oscillators with identical properties are
symmetrically coupled without noise, the theory can also be extended to include heterogeneity
of the oscillators or noise.

The linear coupling with time delay or linear filtering can be realized without measuring the phase values of the oscillator, once the correlation functions of the PSF and the limit-cycle orbit (or their derivatives) are obtained.
The nonlinear coupling requires the measurement of the phase values of the oscillators, but can further improve the linear stability of the synchronized state. We have shown that a simple approximate evaluation of the phase values by a linear interpolation gives reasonable results even though it may yield a somewhat incorrect evaluation of the true phase values.

It is interesting to compare the present analysis for stable synchronization between the two oscillators with the optimization of driving signals for injection locking of a single oscillator, which has been analyzed by Zlotnik {\it et al.}~\cite{zlotnik13} and others (briefly explained in the Appendix B for a simple case).
In Sec. IV C, we have obtained the optimal driving functional. In particular, when ${\rm A}(\psi) = {\rm K}$, where ${\rm K}$ is a constant matrix, the optimal driving signal is
\begin{align}
{\bm G}(\psi) = \frac{1}{2\lambda} {\rm K}^{\dag} {\bm Z}'(\psi)
\end{align}
and the maximum stability is 
\begin{align}
- \Gamma'(0)
&=
\sqrt{ P \la \left| {\rm K}^{\dag} {\bm Z}'(\psi)  \right|^2 \ra_\psi }.
\end{align}
This result coincides with the optimal injection signal for stable synchronization of a single oscillator, obtained by Zlotnik {\it et al.}~\cite{zlotnik13}.
Thus, the optimal coupling between the oscillators is realized by measuring the present phase $\psi$ of the other oscillator and applying a driving signal that is proportional to ${\rm K}^{\dag} {\bm Z}'(\psi)$ to the oscillator.

It is also interesting to note that we have obtained similar expressions for the maximum stability in all examples,
$-\Gamma'(0) = \sqrt{ P \la \cdots \ra_\psi^2 }$,
where $\cdots$ depends on the quantity to be optimized.
This is because we are essentially maximizing the inner product of 
the PSF with the derivative of the driving
signal under a mean-square constraint on the parameters or functions
included in the driving signal in all cases.

The linear coupling schemes in Sec.~III would be easy to realize experimentally.
The nonlinear coupling schemes in Sec.~IV require the evaluation of the phase values
from the oscillators, but can yield an even higher linear stability.
These methods may be useful when higher stability of the in-phase synchronized state
between oscillators is desirable in technical applications.
It would also be interesting to study interactions between rhythmic elements e.g. in
biological systems from the viewpoint of synchronization efficiency.

\acknowledgements

This work is financially supported by JSPS KAKENHI Grant Numbers JP16K13847, JP17H03279, 18K03471, JP18H032, and 18H06478.

\appendix

\section{Models with self coupling}

We here show that the inclusion of self-coupling terms in the model does not alter the linear stability of the in-phase synchronized state under the phase-reduction approximation. 
Suppose that we have additional self-coupling terms in the model as
\begin{align}
\dot{\bm X}_1 =& {\bm F}({\bm X}_1) + \epsilon [ \hat{\bm H}\{ {\bm X}_1^{(t)}(\cdot), {\bm X}_2^{(t)}(\cdot) \} +  \hat{\bm I}\{ {\bm X}_1^{(t)}(\cdot) \} ],
\cr
\dot{\bm X}_2 =& {\bm F}({\bm X}_2) + \epsilon [ \hat{\bm H}\{ {\bm X}_2^{(t)}(\cdot), {\bm X}_1^{(t)}(\cdot) \} +  \hat{\bm I}\{ {\bm X}_2^{(t)}(\cdot) \} ],
\end{align}
where  $\hat{\bm I}\{ {\bm X}^{(t)}(\cdot) \}$ is a functional representing self coupling.
A typical example is coupled oscillators with linear diffusive coupling,
\begin{align}
\dot{\bm X}_1 =& {\bm F}({\bm X}_1) + \epsilon ( {\bm X}_2 - {\bm X}_1 ),
\cr
\dot{\bm X}_2 =& {\bm F}({\bm X}_2) + \epsilon ( {\bm X}_1 - {\bm X}_2 ),
\end{align}
where we may take $\hat{\bm H}\{ {\bm X}_1^{(t)}(\cdot), {\bm X}_2^{(t)}(\cdot) \} = {\bm X}_2(t)$ and $\hat{\bm I}\{ {\bm X}_1^{(t)}(\cdot) \} = - {\bm X}_1(t)$.
By phase reduction, we obtain the phase equations
\begin{align}
\dot{\theta}_1 &= \omega + \epsilon {\bm Z}(\theta_1) \cdot [ {\bm H}( \theta_1, \theta_2 ) + {\bm I}(\theta_1) ],
\cr
\dot{\theta}_2 &= \omega + \epsilon {\bm Z}(\theta_2) \cdot [ {\bm H}( \theta_2, \theta_1 ) + {\bm I}(\theta_2) ],
\end{align}
and the phase coupling function 
\begin{align}
\tilde\Gamma(\phi) 
&= \la {\bm Z}(\psi) \cdot [ {\bm H}( \psi, \psi - \phi ) + {\bm I}(\psi) ] \ra_\psi
\cr
&= \Gamma(\phi) + \la {\bm Z}(\psi) \cdot {\bm I}(\psi) \ra_\psi,
\end{align}
where $\Gamma(\phi)$ is the phase coupling function for the case without the self-coupling term and the second term is a constant. Thus, this model also has the in-phase synchronized state as a fixed point and its linear stability is equal to the case without the self-coupling term,
\begin{align}
- \epsilon \Gamma'(0) = - \epsilon \tilde\Gamma'(0).
\end{align}

\section{Optimal signal for injection locking}

We here briefly review Zlotnik {\em et al.}'s result~\cite{zlotnik13} on the optimal driving signal for injection locking for a simple case.
We consider a limit-cycle oscillator driven by a periodic driving signal whose period is the same as the natural period $T$ of the oscillator, 
\begin{align}
\dot{\bm X} = {\bm F}({\bm X}) + \epsilon {\rm K} {\bm f}(t), \quad {\bm f}(t) = {\bm f}(t+T),
\end{align}
where ${\bm X}$ is the oscillator state, ${\bm F}({\bm X})$ represents its dynamics, and $\epsilon {\rm K} {\bm f}(t)$ is a weak periodic driving signal, where $\epsilon$ is a small positive parameter and a constant matrix ${\rm K} \in {\mathbb R}^{N \times N}$ represents which components of ${\bm X}$ are driven by ${\bm f}(t)$.

By phase reduction, we obtain a phase equation
\begin{align}
\dot{\theta} = \omega + {\bm Z}(\theta) \cdot {\rm K} {\bm f}(t)
\end{align}
for the oscillator phase $\theta$, where ${\bm Z}(\theta)$ is the PSF. Defining $\theta - \omega t = \phi$ and averaging over one oscillation period yields
\begin{align}
\dot{\phi} = \Gamma(\phi).
\end{align}
The phase coupling function $\Gamma(\psi)$ is expressed as
\begin{align}
\Gamma(\phi)
&= \frac{1}{2\pi} \int_0^{2\pi} {\bm Z}(\phi+\psi) \cdot {\rm K} {\bm f}(\psi / \omega) d\psi
\cr
&= \frac{1}{2\pi} \int_0^{2\pi} {\bm Z}(\phi+\psi) \cdot {\rm K} \tilde{\bm f}(\psi) d\psi
\cr
&= \la {\bm Z}(\phi+\psi) \cdot {\rm K} \tilde{\bm f}(\psi) \ra_\psi
\end{align}
where we have defined $\tilde{\bm f}(\psi) =  {\bm f}(\psi / \omega)$.

By choosing the origin of the phase of the periodic signal so that $\Gamma(0) = 0$ holds,
the oscillator synchronizes with the periodic at $\phi = 0$, and the linear stability of this synchronized state is given by
\begin{align}
\Gamma'(0)
&= \la {\bm Z}'(\psi) \cdot {\rm K} \tilde{\bm f}(\psi) \ra_\psi
\cr
&= \la {\rm K}^{\dag} {\bm Z}'(\psi) \cdot \tilde{\bm f}(\psi) \ra_\psi.
\end{align}
We constrain the one-period average of $\tilde{\bm f}(\psi)$ as
\begin{align}
\la | \tilde{\bm f}(\psi) |^2 \ra_\psi = P,
\end{align}
and consider an objective function
\begin{align}
S\{\tilde{\bm f}, \lambda\}
&= - \Gamma'(0) + \lambda( | f |^2 - P )
\cr
&= - \la {\rm K}^{\dag} {\bm Z}'(\psi) \cdot \tilde{\bm f}(\psi) \ra_\psi + \lambda \left( \la | \tilde{\bm f}(\psi) |^2 \ra_\psi - P \right),
\cr
\end{align}
where $\lambda$ is a Lagrange multiplier. Extremum conditions are given by
\begin{align}
\frac{\delta S}{\delta \tilde{\bm f}(\psi)} = - \frac{1}{2\pi} {\rm K}^{\dag} {\bm Z}'(\psi) + \frac{\lambda}{\pi} \tilde{\bm f}(\psi) = 0,
\end{align}
\begin{align}
\frac{\partial S}{\partial \lambda} = \la | \tilde{\bm f}(\psi) |^2 \ra_\psi - P = 0.
\end{align}
The optimal driving signal is given by
\begin{align}
\tilde{\bm f}(\psi) = \frac{1}{2\lambda} {\rm K}^{\dag} {\bm Z}'(\psi)
\end{align}
and the constraint is 
\begin{align}
\frac{1}{4 \lambda^2} \la | {\rm K}^{\dag} {\bm Z}'(\psi) |^2 \ra_\psi = P,
\end{align}
which yields
\begin{align}
\lambda = \sqrt{ \frac{1}{4 P} \la | {\rm K}^{\dag} {\bm Z}'(\psi) |^2 \ra_\psi },
\end{align}
where the negative sign should be taken in order that $\Gamma'(0) < 0$. Thus, the optimal driving signal is given by
\begin{align}
\tilde{\bm f}(\psi) =
- \sqrt{\frac{P}{ \la | {\rm K}^{\dag} {\bm Z}'(\psi) |^2 \ra_\psi }} {\rm K}^{\dag} {\bm Z}'(\psi)
\end{align}
and the maximum linear stability is given by
\begin{align}
- \Gamma'(0) &=  -\la {\rm K}^{\dag} {\bm Z}'(\psi) \cdot \tilde{\bm f}(\psi) \ra_\psi
=
\sqrt{ P \la | {\rm K}^{\dag} {\bm Z}'(\psi) |^2 \ra_\psi }.
\end{align}


\end{document}